\documentclass[%
 reprint,
 amsmath,amssymb,
 aps,
]{revtex4-2}
\usepackage[utf8]{inputenc}
\usepackage{xcolor}
\usepackage{eqnarray,amsmath}
\usepackage[title]{appendix}
\usepackage{comment}
\usepackage{graphicx}
\usepackage{dcolumn}
\usepackage{bm}
\usepackage{braket}
\usepackage{comment}

\begin{document}

\preprint{APS/123-QED}

\title{Graphene-based quantum heterospin graphs}

\author{Gabriel Martínez-Carracedo$^{1,2}$}
\author{Amador García-Fuente$^{1,2}$}
\author{László Oroszlány$^{3,4}$}
\author{László Szunyogh$^{5,6}$}
\author{Jaime Ferrer$^{1,2}$}

\affiliation{
$^1$Departamento de Física,  Universidad de Oviedo,  33007 Oviedo, Spain\\
$^2$Centro de Investigación en Nanomateriales y Nanotecnología, Universidad de Oviedo-CSIC, 33940, El Entrego, Spain\\
$^3$Department of Physics of Complex Systems, Eötvös  Loránd University, 1117 Budapest, Hungary\\
$^4$Wigner Research Centre for Physics, H-1525, Budapest, Hungary\\
$^5$Department of Theoretical Physics, Institute of Physics, Budapest University of Technology and Economics, 
M\H{u}egyetem rkp. 3., H-1111 Budapest, Hungary \\
$^6$HUN-REN-BME Condensed Matter Research Group, Budapest University of Technology and Economics, 
M\H{u}egyetem rkp. 3., H-1111 Budapest, Hungary
}

\begin{abstract}
We investigate from first principles a variety of low-dimensional open quantum spin systems based on magnetic nanographene structures that contain spin-1/2 and spin-1 triangulenes and/or olympicenes. These graphene nanostructures behave as localized spins and can be effectively described by a quantum bilinear-biquadratic Heisenberg Hamiltonian, for which we will compute the energy spectrum and the quantum numbers associated with the low-energy eigenstates. We propose the experimental realization of antiferromagnetic alternating spin chains using these graphene nanostructures, which result in ferrimagnetic systems whose ground state spin and degeneracy depend on the length of the chain. We identify a double degeneracy in the total spin quantum number $S$ of the first excited state in three-leg spin graphs (3-LSGs) and other heterospin nanostructures, which depends on both the number of sites and the spin species, and originates from the swapping transformation symmetry of the Hamiltonian. Numerical simulations indicate that this degeneracy remains largely robust for $N=7$ spin-1 3-LSGs under realistic perturbations present in experimental conditions.

\end{abstract}


\maketitle


\section{Introduction}
\begin{figure}[h]
    \centering
    \includegraphics[width=\columnwidth]{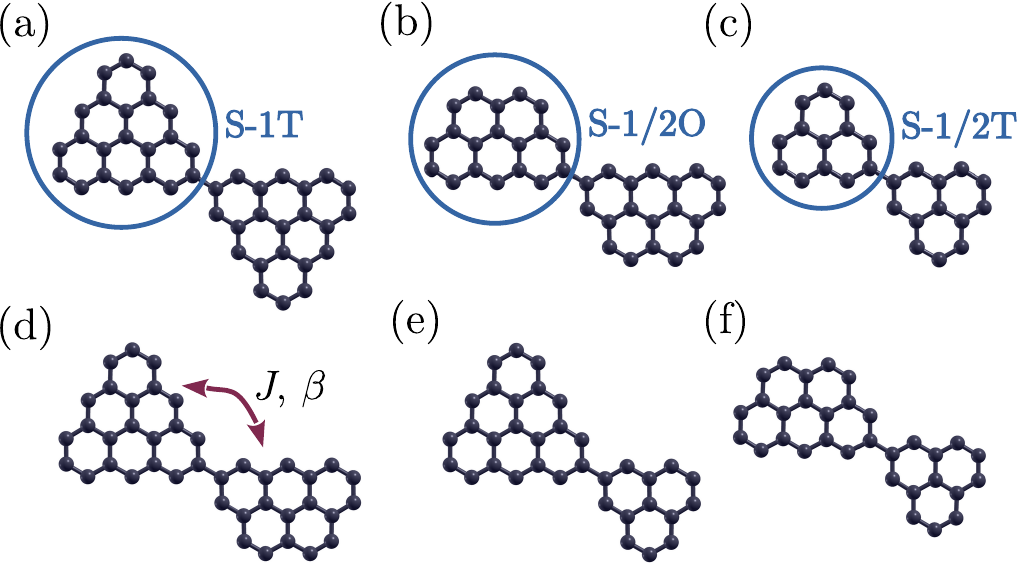}
    \caption{(a-f) Magnetic dimer systems coupled by bilinear $J$ and biquadratic $\beta$ exchange constants. These dimers are composed of different graphene-based MBBs: S-1T, S-1/2T, and S-1/2O. The dimers are: (a) S-1T and S-1T, (b) S-1/2O and S-1/2O, (c) S-1/2T and S-1/2T, (d) S-1T and S-1/2O, (e) S-1T and S-1/2T, and (f) S-1/2O and S-1/2T.}
    
    \label{junctions}
\end{figure}

Harnessing quantum-regime functionalities without relying on rare-earth elements represents a significant step toward the development of globally-sourced and more accessible quantum technologies. Moreover, implementing using widely accessible materials, such as graphene-based devices \cite{Garreis2024, AlonsoCalafell2019, Zhang2023, Oh2021} or other molecular systems \cite{Lehmann2009}, has garnered significant attention from both theorists and experimentalists. Quantum spin systems provide an excellent playground for exploring the physical properties of strongly correlated systems \cite{PhysRevX.11.011011,PhysRevLett.121.037204} and they can be used to construct platforms for quantum computing \cite{PRXQuantum.3.020343}. Some examples of spin quantum systems include two-dimensional magnets \cite{maggenom,Huang2017,Boix-Constant2025,PhysRevB.110.184406}, quantum dots \cite{RevModPhys.79.1217,Appel2025} or topological spin textures \cite{Zuo2021,Tejo2021}. In this paper, we focus on specific magnetic nanographene structures \cite{Fujii2013,Cheng2022,Song2024,Magda2014} and use them as magnetic building blocks (MBBs) to construct various interacting spin systems. This is because these MBBs possess a magnetic moment whose magnitude is determined by Lieb's theorem \cite{PhysRevLett.62.1201}, taking into account the imbalance of carbon atoms in each sublattice. Several graphene-based MBBs have recently attracted extensive attention as platforms for spin clusters and spin chains \cite{10.1103/PhysRevResearch.5.043226,Fu2025,10.1103/PhysRevResearch.6.043262,deOteyza2025,Zhao2024}. We  focus on three MBBs that have already been synthesized experimentally: the spin-1 triangulene (S-1T) \cite{Mishra2019}, the spin-1/2 olympicene (S-1/2O) \cite{exp_olimp}, and the spin-1/2 triangulene (S-1/2T) \cite{exp_spin12}, which are circled in blue in Figs. \ref{junctions}(a-c). Since their magnetic moments are sufficiently localized and fixed, and there is no source of anisotropy, the interaction between MBBs can be appropriately described by an isotropic quantum Heisenberg Hamiltonian. It has been shown that to describe effectively the exchange interaction between MBBs in these graphene nanostructures  a biquadratic interaction  \cite{exp_spin1,PhysRevB.108.155423} must be included. The bilinear-biquadratic (BLBQ) Hamiltonian  reads
\begin{equation}
\hat{H}_{\text{BLBQ}}=\dfrac{1}{2}\sum\limits_{i\neq j}J_{ij}\left(\hat{{\bf S}}_i\cdot\hat{{\bf S}}_j+\beta_{ij}\left(\hat{{\bf S}}_i\cdot\hat{{\bf S}}_j\right)^2\right)
    \label{H_heisenberg}
\end{equation}
where $\hat{{\bf S}}_i = (\hat{{ S}}_i^x, \hat{{ S}}_i^y, \hat{{ S}}_i^z)$ is a vector whose components are the spin-$S$ operators in the $x$, $y$, and $z$ directions for the $i$th MBB. Note that for each site $i$, the spin operator may differ (either spin-1 or spin-1/2) depending on the type of MMB associated with $i$. $J_{ij}$ are the isotropic exchange constants, and $\beta$ are parameters that modulate the biquadratic exchange interactions being a higher-order term beyond the bilinear exchange. \\
 In this paper, we present an {\sc ab-initio} study that combines the exact diagonalization (ED) method \cite{PhysRevLett.113.127204,PhysRevB.103.205122} with density functional theory (DFT) and the LKAG formalism to extract information about the energy gaps and state degeneracies of multi-legged spin systems described by Eq. (\ref{H_heisenberg}). The structure of the paper is as follows. In Section \ref{Sec_method}, we explain the method used to characterize the nanographene spin structures. In Section \ref{secresults}, we present the results for alternating spin chains, three-leg spin graphs (3-LSGs) and some multi-leg and mixed-spin graphs. In this section, we also discuss and characterize the degeneracy that arises in 3-LSGs when  a swapping symmetry group isomorphic to the $C_{3v}$ point group is present. We also perform numerical simulations that indicate this degeneracy remains robust under realistic perturbations present in experimental conditions. Finally, we summarize the results and present our conclusions in Section \ref{Sec_conclusions}. 
 
\section{Computational method}\label{Sec_method}
We follow a three-step procedure to study and characterize these magnetic graphene-based nanostructures. The first step is to compute the electronic Hamiltonian for the various dimer systems depicted in Fig. \ref{junctions} using Density Functional Theory (DFT) \cite{PhysRev.136.B864,PhysRev.140.A1133} via the {\sc SIESTA} code \cite{Soler2002}. For the {\sc SIESTA} calculations, we employed the Generalized Gradient Approximation (GGA) within the Perdew-Burke-Ernzerhof (PBE) scheme \cite{PhysRevLett.77.3865}. We used a mesh cutoff of 500 Ry for real-space integrals and a double-$\zeta$ polarized basis to ensure good convergence. Atomic forces were relaxed to a tolerance of 0.005 eV/\AA\, using molecular dynamics. The edges of these dimer systems were passivated with hydrogen atoms. Mulliken analysis confirmed that the total spin of S-1T is equal to $\hbar$, while the total spins of S-1/2T and S-1/2O are $\hbar/2$.\\

\begin{table}[ht!]
\small
  \caption{Bilinear and biquadratic exchange constants for the dimer systems depicted in Fig. \ref{junctions}.}
  \begin{tabular*}{0.5\textwidth}{@{\extracolsep{\fill}}ccc}
    \hline
    Dimer Coupling & $J$ (meV) & $\beta$ \\
    \hline
     (a) & 15.7 & 0.02\\
     (b) & 38.4& 0.08\\
     (c)& 48.8&0.08 \\
     (d)& 24.4& 0.04 \\
     (e)&27.5& 0.04 \\
     (f)& 43.2 &0.08 \\  
    \hline
  \end{tabular*}
  \label{table_coupling}
\end{table}
The second step is to compute the bilinear and biquadratic exchange constants of the Hamiltonian in Eq. (\ref{H_heisenberg}) in the context of the the magnetic force theorem and the LKAG formalism \cite{Liechtenstein1987,PhysRevB.68.104436,PhysRevB.99.224412} using the electronic Hamiltonian previously computed as input, as explained in more detail in Ref. \cite{PhysRevB.107.035432}. These exchange parameters are obtained by mapping to the classical version of Eq. (\ref{H_heisenberg}), where spin operators are replaced by unit vectors. Therefore, Table \ref{table_coupling} compiles the exchange parameters for the different dimer systems shown in Fig. \ref{junctions}, excluding the $S$ factors. Notice that all isotropic exchange couplings favor antiferromagnetic (AFM) alignments between the two spins. Throughout this work, we assume exchange interactions between the MBBs exist only among nearest neighbors \cite{PhysRevB.107.035432,Saleem2024,exp_spin1}.  Moreover, to avoid undesired contributions to the exchange parameters, we have projected the magnetic entities onto the $2p_z$-orbitals \cite{PhysRevB.108.214418}, which are responsible for the magnetism in these zigzag-edged nanographene structures \cite{PhysRevLett.99.177204,Ominato2013}. The reported values for the exchange coupling $J$ between two S-1/2O \cite{exp_olimp}, two S-1/2T \cite{exp_spin12}, and two S-1T \cite{exp_spin1} are 38, 45, and 18 meV, respectively. These values are in good agreement with the theoretical results we report in Table \ref{table_coupling}(a-c). Note that for interacting spin-1–1/2 and spin-1/2–1/2 pairs described by Eq. (\ref{H_heisenberg}), an effective $J^{\text{eff}}_{ij}$ combining the bilinear and biquadratic interactions could, in principle, be introduced. However, our computational method is based on a mapping to the classical form of Eq. (\ref{H_heisenberg}), which allows us to distinguish explicitly between the bilinear and the higher-order biquadratic contributions. \\

Lastly, the third step is to perform Exact Diagonalization (ED) on the Hamiltonian $\hat{H}_{\text{BLBQ}}$ given by Eq. (\ref{H_heisenberg}) using the exchange parameters previously obtained.  Since the operators $\hat{H}_{\text{BLBQ}}$, $\hat{S}^2_T=(\sum\limits_i\hat{{\bf S}}_i)^2$ and $\hat{S}^z_T=\sum\limits_i\hat{{ S}}_i^z$ commute with each other, the eigenstates of $\hat{H}_{\text{BLBQ}}$ labelled by $\lambda$ can be characterized with the respective quantum numbers $S$ and $M$ and are computed using ED. As we will see in section \ref{3legsection}, for certain eigenstates a degeneracy not related to these quantum numbers emerges that is related to an additional symmetry of  $\hat{H}_{\text{BLBQ}}$.

\section{Results}
\label{secresults}
\subsection{Alternating spin-1 and spin-1/2 chains}
The theory of alternating quantum spin chains has been studied for more than 60 years \cite{LiebJ,PhysRevB.55.8894,SBrehmer_1997,PhysRevB.64.134408,WU20111428}. In their seminal work, Lieb and Mattis \cite{LiebJ} presented significant analytical results for bipartite spin systems described by a quantum AFM isotropic Heisenberg Hamiltonian. In the particular case of bipartite chains, the spin species in each sublattice may not be equal; for example, one sublattice may consist of spin-1 and the other of spin-1/2. When this occurs, we refer to it as an alternating spin chain. The Lieb-Mattis theorem \cite{LiebJ} can be applied to an alternating AFM spin chain as follows. Suppose that the exchange interaction occurs only between the first nearest neighbors, and let $\mathcal{S}_A$ ($\mathcal{S}_B$) denote the maximum possible spin on sublattice $A$ ($B$). According to the theorem, the quantum spin number $S$ for the ground state (GS) of the system is given by $S = |\mathcal{S}_A - \mathcal{S}_B|$. For an alternating spin-1 and spin-1/2 of length $N$ the spin of the GS will be given by 
\begin{equation}
S=|\left \lfloor{N/2}\right \rfloor+\text{mod}(N,2)-0.5\left \lfloor{N/2}\right \rfloor|,
    \label{SofN}
\end{equation}
where $\left \lfloor{\cdot}\right \rfloor$ is the floor function. This reflects the net magnetization as a function of $N$ and the ferrimagnetic behavior, as $S$ grows linearly with $N$. The degeneracy $g = 2S + 1$ of the ground state is also linearly dependent on $N$. It has been shown that the ground state differs from the ferrimagnetic Néel state due to quantum fluctuations \cite{SBrehmer_1997}.\\

\begin{figure}[htb]
    \centering
    \includegraphics[width=\columnwidth]{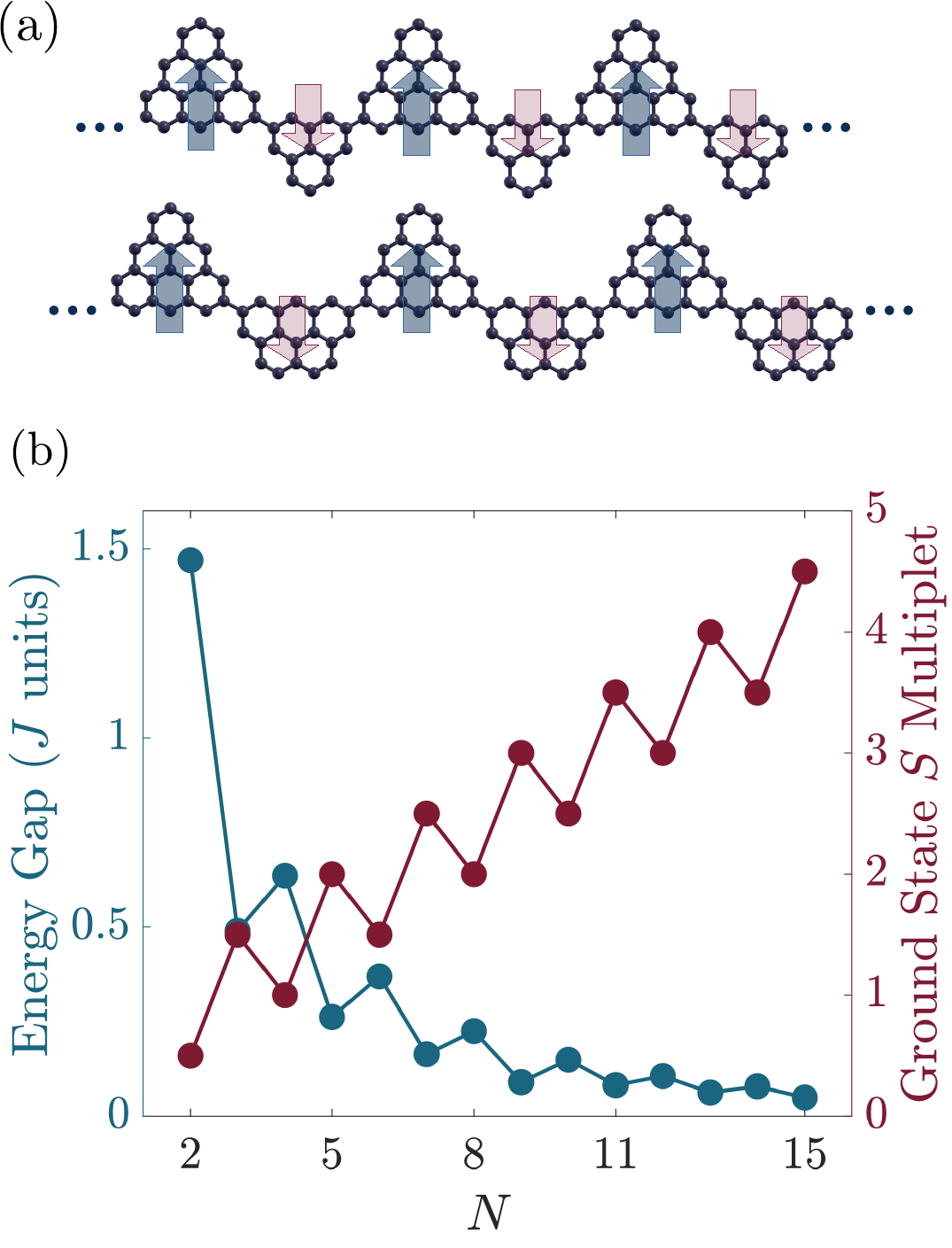}
    \caption{(a) shows the two types of ferrimagnetic alternating spin-1 and spin-1/2 chains that can be formed from different graphene MBBs. The top chain is made from S-1T and S-1/2T MBBs, while the bottom chain is made from S-1T and S-1/2O MBBs. Blue (red) arrows represent spin-1 (spin-1/2) sites. (b) shows the energy gap between the GS and the first excited state in blue, and the quantum spin number $S$ for the GS in red, as a function of the chain length $N$. Both calculations were performed using ED over Eq. (\ref{H_heisenberg_chain}) taking $\beta$=0.04.}
    
    \label{alternatingchain}
\end{figure}

In this paper, we propose to use MBBs such as S-1T, S-1/2T or S-1/2O to construct an alternating quantum spin chain as depicted in Fig. \ref{alternatingchain}(a). These chains are described by a simplified form of Eq. (\ref{H_heisenberg}) 
\begin{equation}
\hat{H}_{\text{chain}}=\sum\limits_{i}J\left(\hat{{\bf S}}_i\cdot\hat{{\bf S}}_{i+1}+\beta\left(\hat{{\bf S}}_i\cdot\hat{{\bf S}}_{i+1}\right)^2\right).
    \label{H_heisenberg_chain}
\end{equation}

Based on the results mentioned above, we show that by engineering such a system, the ground state exhibits ferrimagnetic behavior, with both the spin number $S$ and the degeneracy of the GS depending on the length of the chain. At the top of Fig. \ref{alternatingchain}(a), we depict an alternating spin-1 and spin-1/2 chain composed of S-1T and S-1/2T, while at the bottom, the chain is composed of S-1T and S-1/2O. Different types of alternating spin chains can be constructed using different MBBs, such as aza-triangulenes (\cite{Wang2022,Vilas-Varela2023}) or larger triangulenes with higher spin moment. The isotropic and biquadratic exchange constants $J$ and $\beta$ for these chains are given in Table \ref{table_coupling}. In Fig. \ref{alternatingchain}(b), we show the energy gap to the first excited state as a function of $N$ in blue. For the alternating spin chains displayed in Fig. \ref{alternatingchain}(b), the energy gap is on the order of a few meV, which lies within the resolution range of scanning tunneling microscopy (STM) experiments \cite{exp_spin1,Lawrence2020}. In the limit of infinite chain length, the energy gap vanishes \cite{PhysRevB.55.8894}. The total spin $S$ of the GS is shown in red as a function of $N$, confirming that it follows Lieb's theorem, as given in Eq. (\ref{SofN}).

\subsection{Three-Leg Spin Graphs}\label{3legsection}

\begin{figure*}[t]
    \centering
    \includegraphics[width=2\columnwidth]{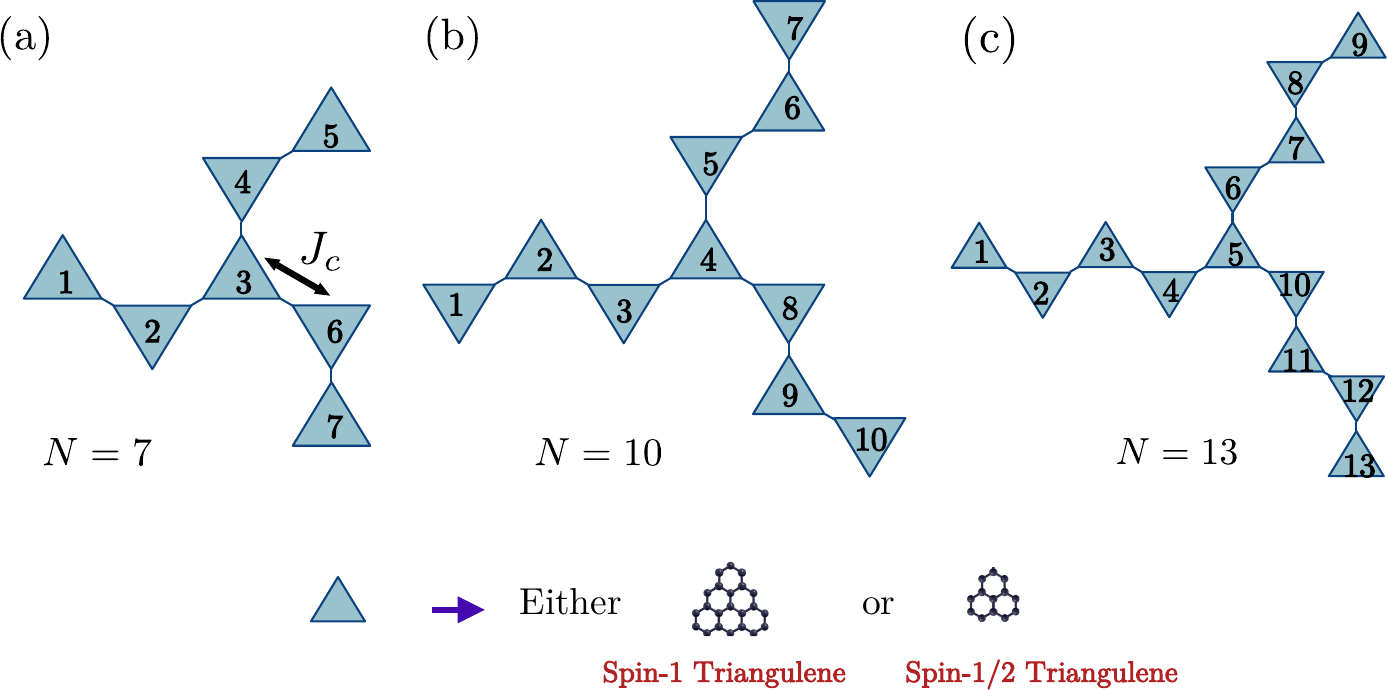}
    \caption{(a-c) 3-LSGs of lengths 
$N=7$, 10, and 13 MBBs, respectively. Each blue triangle either represents a spin-1 or spin-1/2 MBB (S-1T or S-1/2T), coupled via isotropic and biquadratic exchange interactions between nearest neighbors (see Table \ref{table_coupling}). Site labels are shown in black for the $N=7$, 10, and 13 3-LSGs. For $N=7$, $J_c$ denotes the isotropic exchange coupling between sites 3 and 6.}
    
    \label{3legchain1}
\end{figure*}

When a spin system cannot be divided into sublattices, as in the case of 3-LSG, Lieb's theorem \cite{LiebJ} is no longer applicable. To the best of our knowledge, there is no theorem or result that could predict the ground state of the Hamiltonian given by Eq. (\ref{H_heisenberg}) as a function of the couplings and spin constituents.\\
 We will devote this section to study 3-LSG whose MBBs are either S-1T or S-1/2T, respectively. We will perform ED on Eq. (\ref{H_heisenberg}) for three different 3-LSG lengths, $N$=7, 10, and 13 spin-1 and spin-1/2 sites, as depicted in Fig. \ref{3legchain1}. In Table \ref{table_multiplets}, we compile the total spin quantum number for the GS and the first excited state (ES), as well as the energy gap between them, for the six 3-LSGs. We find that the first ES is two-fold degenerate in energy and spin $S$ quantum number. We will refer to this as double-$S$ degeneracy. For example, for a $N=10$ 3-LSG made by spin-1 MBBs, the first ES is composed by two-fold degenerate $S=1$ states which gives a total of degenerate six states (two $M=-1$, two $M=0$ and two $M=1$ states). Moreover, following this case, in the presence of a magnetic field represented by the Zeeman term, $H_Z=-B\hat{S}_T^z$, the ES will have $M=1$ while remaining two-fold degenerate.\\ This phenomenon is common to all six 3-LSGs studied in this section for different MBBs and is robust for different $\beta$ values in the range [0, 0.1] within the same phase. 
 For larger values of $\beta$ in 3-LSGs composed of spin-1 MBBs, the energy spectrum changes significantly, with the quantum numbers and degeneracies of the GS and first ES differing from those observed at smaller $\beta$ values.
 To the best of our knowledge, the degeneracy in the $S$ and $M$ quantum numbers has not been characterized for these 3-LSG systems in the literature. Indeed, eigenstates of $\hat{H}$ cannot be unambiguously classified by $S$ and $M$. We address this issue in more detail later in this section.\\
 
 \begin{table}[htb!]
\small
  \caption{$S$ value for the GS and the two-fold degenerate first ES, computed via ED using the Hamiltonian in Eq. (\ref{H_heisenberg}), for a 3-LSG composed of either spin-1 or spin-1/2 MBBs of varying lengths ($N=7$, 10 and 13). The energy gap between the GS and ES is also provided.}
\begin{tabular*}{0.5\textwidth}{@{\extracolsep{\fill}}cccc}
    \hline
    3-LSG & GS &  ES & Gap ($J$ units) \\ 
    \hline
     $N=7$, spin-1/2 & $S=1/2$ & $S=1/2$ & 0.72\\ 
     $N=7$, spin-1&$S=1$ & $S=0$ &0.60 \\ 
     $N=10$, spin-1/2& $S=1$& $S=0$ &0.27 \\ 
     $N=10$, spin-1& $S=2$& $S=1$ &0.40 \\ 
     $N=13$, spin-1/2& $S=1/2$ &$S=1/2$ & 0.46\\ 
     $N=13$, spin-1& $S=1$&$S=0$ & 0.27\\ 
    \hline
\end{tabular*}
  \label{table_multiplets}
\end{table}

\begin{figure*}[htb!]
    \centering
    \includegraphics[width=2\columnwidth]{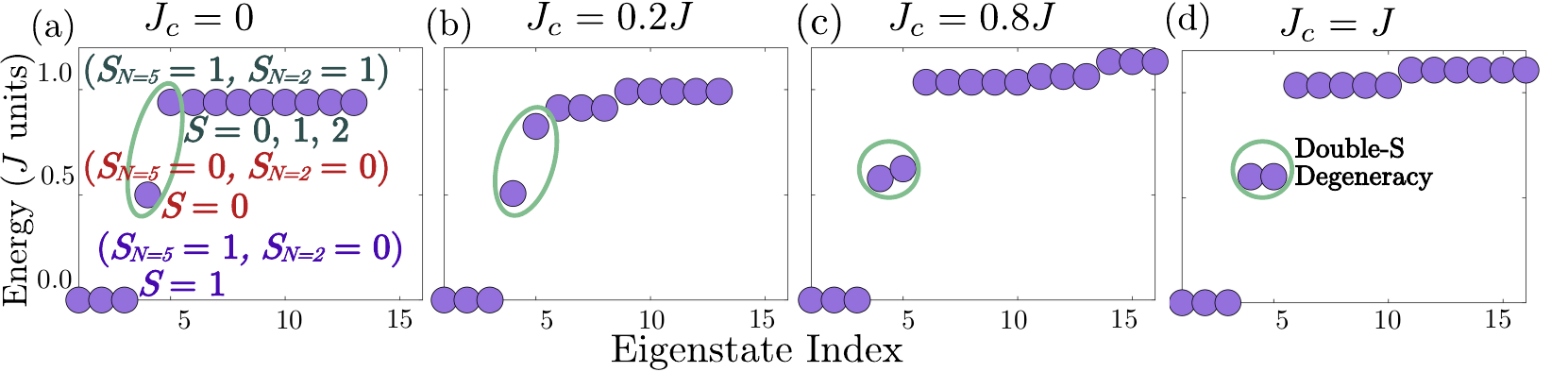}
    \caption{(a–d) ED calculations based on the Hamiltonian given by Eq. (\ref{H_heisenberg}) for an $N=7$ three-leg spin graph composed of spin-1 MBBs. The exchange coupling $J_c$ connects sites three and six, following the labeling in Fig. \ref{3legchain1} (a), thereby coupling two chains: one with $N=2$ and the other with $N=5$. As $J_c$ increases, the two singlet states circled in green - one in the first ES and the other in the second ES - progressively approach each other in energy until they become degenerate at $J_c=J$.}
    \label{3legchain2}
\end{figure*}

 To understand how the double-$S$ degeneracy arises, we focus on the $N = 7$ 3-LSG composed of spin-1 MBBs, where the isotropic exchange coupling $J_c$ connects sites 3 and 6, as shown in Fig. \ref{3legchain1}(a). When $J_c=0$ the system consists of two uncoupled spin-1 chains: one of length $N=5$ and the other of length $N=2$. The GS of an open spin-1 chain depends on the number of sites \cite{Haldane1983,PhysRevLett.59.799,PhysRevB.107.035432,Jaworowski2017}, being a singlet (triplet) if the number of sites is even (odd). Thus, for the $N=2$ chain, the GS and ES are a singlet ($S_{N=2}=0$) and triplet ($S_{N=2}=1$), respectively, while for the $N=5$ chain, the GS and ES are reversed, being a triplet ($S_{N=5}=1$) and singlet ($S_{N=5}=0$), respectively. Fig. \ref{3legchain2}(a) illustrates how the eigenvalues of the Hilbert space, formed by the $N=2$ and $N=5$ chains, are combined through the tensor product. The GS is a triplet, arising from $S_{N=5}=1$ and $S_{N=2}=0$, while the first ES is a singlet, formed by 
$S_{N=5}=0$ and $S_{N=2}=0$. The second ES consists of a singlet, a triplet, and a quintuplet, resulting from the combination of the two triplets, 
$S_{N=5}=1$ and $S_{N=2}=1$. Figs. \ref{3legchain2}(b–d) illustrate how the singlet, formed by the combination of 
$S_{N=5}=1$ and $S_{N=2}=1$, decreases in energy as the exchange coupling $J_c$ increases, approaching the singlet ES. The energy difference between the first and second singlet ESs for 
$J_c=0.2J$ ($J_c=0.8J$) is approximately 40\% (8\%). When $J_c=J$, both singlets become degenerate giving rise to the double-$S$ degeneracy. The degeneracy mechanism shown in Figs. \ref{3legchain2}(a–d) is valid only for the $N=7$ 3-LSG with spin-1 MBBs. This decoupling approach is analogous for the 3-LSGs shown in Table \ref{table_multiplets}. As in all cases, the system separates into two chains in the same manner as discussed above. As we will show below, the degeneracy of the first ES originates from the $C_{3v}$ symmetry. In section \ref{appJK}, we will discuss how possible perturbations in an experimental environment may affect this degeneracy.\\

 As what follows we show that the Hamilton operator of three-leg spin graphs exhibits a symmetry that leads to double degeneracies of some ESs. 
We first introduce the swap operator, which has been widely used in quantum computing \cite{Siewert2022,Nielsen2010}, but not in theoretical magnetism. 
Suppose an operator $\hat{A}$ composed by a tensorial product of three operators $\hat{A}_1$, $\hat{A}_2$ and $\hat{A}_3$ acting on different Hilbert spaces $\mathcal{H}_1$, $\mathcal{H}_2$ and $\mathcal{H}_3$ spanned by  $\{\ket{v_1}\dots\ket{v_{N_v}}\}$, $\{\ket{u_1}\dots\ket{u_{N_u}}\}$ and $\{\ket{w_1}\dots\ket{w_{N_w}}\}$, respectively. This reads 
\begin{equation}
    \begin{split}
\hat{A}&=\hat{A}_1\otimes\hat{A}_2\otimes\hat{A}_3=\\&=\sum A^1_{v\tilde{v}}A^2_{u\tilde{u}}A^3_{w\tilde{w}}\ket{v}\bra{\tilde{v}}\otimes\ket{u}\bra{\tilde{u}}\otimes\ket{w}\bra{\tilde{w}}
    \end{split}
    \label{a1a2a3}
\end{equation}
where we omitted  to present the indices in the sums for simplicity. The unitary operator that swaps the order in the tensor product of the second and third operators in Eq. \eqref{a1a2a3} is given by

\begin{equation}
\hat{U}_{23}=\hat{\mathcal{I}}_1\otimes\sum\left(\ket{\tilde{u}}\bra{\tilde{w}}\otimes\ket{w}\bra{u}\right)
    \label{swap23}
\end{equation}

where $\hat{\mathcal{I}}_1$ is the identity operator acting on $\mathcal{H}_1$. It can be readily proven that $\hat{U}_{23}\,\hat{A}\,\hat{U}_{23}^\dagger=\hat{A}_1\otimes\hat{A}_3\otimes\hat{A}_2$. A general form of the swap operator that exchanges $i$th and $j$th spaces in 
$\mathcal{H}_1\otimes\mathcal{H}_2\otimes\cdots\otimes\mathcal{H}_N$ is given by the following expression
\begin{equation}
\hat{U}_{ij}=\sum\limits_{\alpha\beta}\hat{\mathcal{I}}_1\otimes\hat{\mathcal{I}}_2\otimes\cdots\otimes\underset{i\text{th site}}{\ket{i_\alpha}\bra{j_\beta}}\otimes\cdots\otimes\underset{j\text{th site}}{\ket{j_\beta}\bra{i_\alpha}}\otimes\cdots\otimes\hat{\mathcal{I}}_N
    \label{swapij}
\end{equation}
where  $\{\ket{i_1},\cdots,\ket{i_{N_i}}\}$ and $\{\ket{j_1},\cdots,\ket{j_{N_j}}\}$ span $\mathcal{H}_i$ and $\mathcal{H}_j$, respectively.
It can be immediately seen that $\hat{U}_{ij}$ is unitary $\hat{U}_{ij}\hat{U}_{ij}^\dagger=\hat{U}_{ij}^\dagger\hat{U}_{ij}=\hat{\mathcal{I}}_1\otimes\cdots\otimes\hat{\mathcal{I}}_i\otimes\cdots\otimes\hat{\mathcal{I}}_j\otimes\cdots\otimes\hat{\mathcal{I}}_N$.\\

 We have realized that 3-LSGs possess swapping transformations
that leave $\hat{H}_{\text{BLBQ}}$ invariant. For example, the operator $\hat{U}^{C^+}_N$ that generates an cyclic swapping transformation, namely, an anticlockwise rotation of the legs on a 3-LSG of $N$ sites can be constructed as a product of the swapping operators given in Eq.~\eqref{swapij}: 
\begin{eqnarray}
&\hat{U}^{C^+}_{N=7}=\hat{U}_{6,4}\hat{U}_{2,6}\hat{U}_{7,5}\hat{U}_{1,7} \label{C3N7}\\
&\hat{U}^{C^+}_{N=10}=\hat{U}_{5,8}\hat{U}_{3,5}\hat{U}_{6,9}\hat{U}_{2,6}\hat{U}_{7,10}\hat{U}_{1,7} \label{C3N10}\\
&\hat{U}^{C^+}_{N=13}=\hat{U}_{6,10}\hat{U}_{4,6}\hat{U}_{7,11}\hat{U}_{3,7}\hat{U}_{8,12}\hat{U}_{2,8}\hat{U}_{9,13}\hat{U}_{1,9}.
\label{C3N13}
\end{eqnarray}
where the labeling of the sites in the 3-LSG is described in Fig. \ref{3legchain1}. 
$\hat{U}^{C^+}_N$ is a unitary operator that commutes with $\hat{H}_{\text{BLBQ}}$, $\hat{S}^2_T$ and $\hat{S}^z_T$. Clearly, the adjoint of $\hat{U}^{C^+}_N$, i.e., the clockwise rotation $\hat{U}^{C^-}_N$, also commutes with the Hamiltonian. Moreover, the combination of swapping operators that interchange two legs in 3-LSGs also leaves the Hamiltonian invariant. We will denote them by $\hat{U}^{\sigma_I}_N$, where $I=1,\,2,\,3$ labels the leg which is unaffected by the symmetry transformation. 

\begin{figure}[h]
    \centering
    \includegraphics[width=\columnwidth]{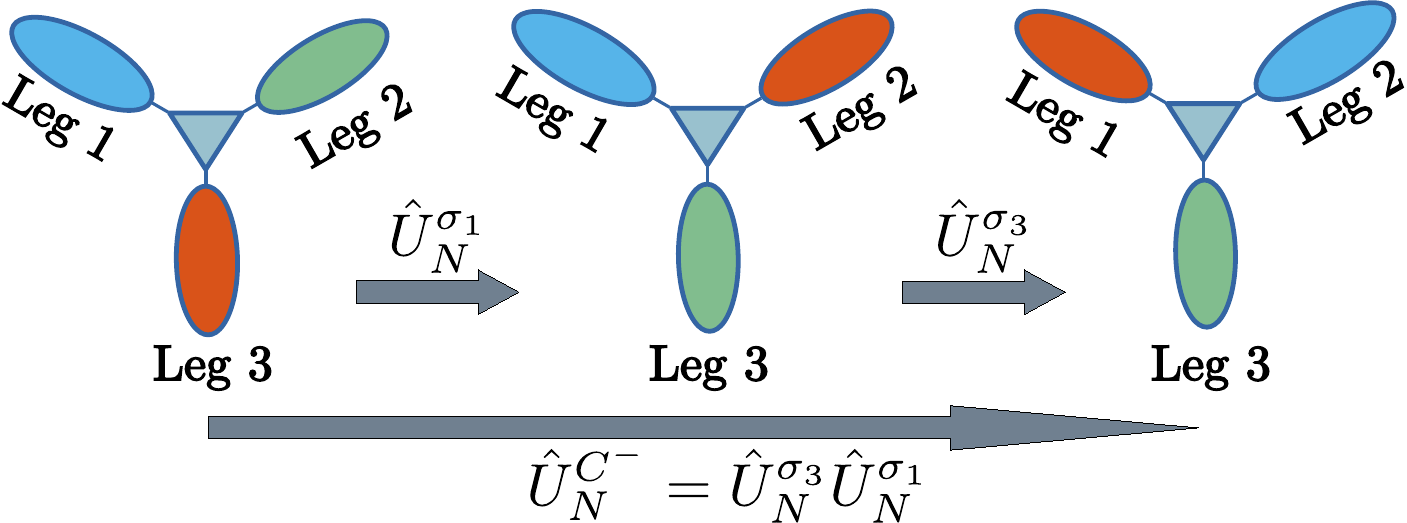}
    \caption{ The construction of the cyclic swapping transformation $\hat{U}^{C^-}_{N}$ for a 3-LSG is illustrated as a subsequent action of the transformation  $\hat{U}^{\sigma_3}_N$ and $\hat{U}^{\sigma_1}_N$. Colors indicate the different spin operators associated with the legs. }
    
    \label{schemeU}
\end{figure}

Consequently, there exist six swapping symmetry transformations $\{\hat{U}^{I}_N$, $\hat{U}^{C^+}_N$, $\hat{U}^{C^-}_N$, $\hat{U}^{\sigma_1}_N$, $\hat{U}^{\sigma_2}_N$, $\hat{U}^{\sigma_3}_N\}$, where  $\hat{U}^{I}_N$ stands for the identity transformation. These transformations form a group, which is isomorphic to the $C_{3v}$ point group. The multiplication table of the group elements can easily be checked, and, as an example, a visual representation of the product  $\hat{U}^{\sigma_3}_N \hat{U}^{\sigma_1}_N = \hat{U}^{C^-}_N$ is shown in Fig.~\ref{schemeU}.

The swapping symmetry group has then two one-dimensional and one two-dimensional irreducible representations, $A_1$, $A_2$ and $E$, respectively. As well-known from representation theory, the eigenstates of the Hamiltonian can be classified according to the irreducible representations.  
For the case $N=7$ 3-LSG of spin-1, we numerically checked that the three eigenstates with lowest energy correspond to the one-dimensional total symmetric $A_1$ representation, i.e., they remain invariant upon any swapping symmetry transformation. The degeneracy of these states is then clearly due to the $M=0,\,\pm 1$ quantum numbers of the triplet ($S=1$) spin-state. 
On the contrary, by applying the swapping transformations the two states of the first ES transformed  into their linear combinations. Though the matrix representations of the swapping transformations may depend on the numerical procedure the ED is performed and also on the parameter $\beta$, it is always possible to find a unitary transformation of these two states such that the representation of the symmetry transformations recovered the well-established two-by-two matrices of the $E$ irreducible representation \cite{ADresselhaus}:
\begin{equation}
\begin{split}
&D(\hat{U}^{I}_N)=\begin{pmatrix}
1 & 0 \\
0 & 1
\end{pmatrix},\, D(\hat{U}^{C^+}_N)=\begin{pmatrix}
-\tfrac{1}{2} & -\tfrac{\sqrt{3}}{2} \\
\tfrac{\sqrt{3}}{2} & -\tfrac{1}{2}
\end{pmatrix},\\& D(\hat{U}^{C^-}_N)=\begin{pmatrix}
-\tfrac{1}{2} & \tfrac{\sqrt{3}}{2} \\
-\tfrac{\sqrt{3}}{2} & -\tfrac{1}{2}
\end{pmatrix},\, D(\hat{U}^{\sigma_1}_N)=\begin{pmatrix}
-1 & 0 \\
0 & 1
\end{pmatrix},\\ &D(\hat{U}^{\sigma_2}_N)=\begin{pmatrix}
\tfrac{1}{2} & -\tfrac{\sqrt{3}}{2} \\
-\tfrac{\sqrt{3}}{2} & -\tfrac{1}{2}
\end{pmatrix},\, D(\hat{U}^{\sigma_3}_N)=\begin{pmatrix}
\tfrac{1}{2} & \tfrac{\sqrt{3}}{2} \\
\tfrac{\sqrt{3}}{2} & -\tfrac{1}{2}
\end{pmatrix}.
\end{split}
\end{equation}
This unambiguously proves that the two-fold degeneracy of the first ES of 3-LSGs is dictated by the swapping symmetry of the Hamiltonian.  It is also clear that the six-fold degenerate fourth ES with $S=1$ of the $N=7$ 3-LSG, see Fig.~\ref{3legchain2}(d), also corresponds to the $E$ irreducible representation.

\begin{figure}[h]
    \centering
    \includegraphics[width=\columnwidth]{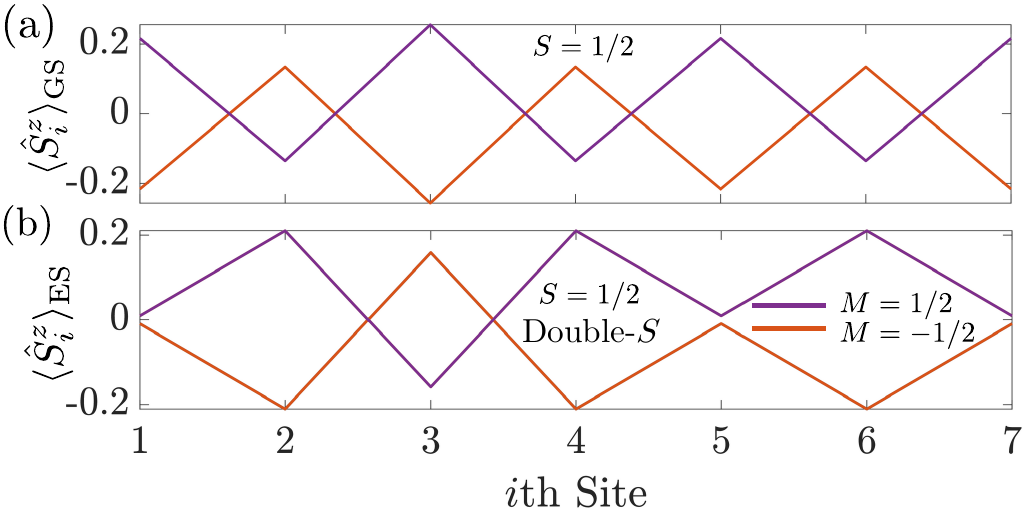}
    \caption{(a) and (b) show the expectation value $\langle\hat{S}_i^z\rangle$ for the GS and the first ES, respectively, for an $N=7$ 3-LSG composed of S-1/2Ts. In (b) the ES is composed by four states $\ket{\lambda_{\text{ES}},S=1/2,M=\pm1/2 }$ because it is double-$S$ degenerate. The labeling of the spin sites is given by Fig. \ref{3legchain1}.}
    
    \label{expectationvalues}
\end{figure}
The expectation values of $\hat{S}_i^z$ for the GS and the first ES are shown in Fig. \ref{expectationvalues} for an $N=7$ 3-LSG of spin-1/2 MBBs. Notice that $\langle\hat{S}_i^z\rangle$ takes the same values for the tuples 1-5-7 and 2-4-6 due to the $C_3$ swapping symmetry given by Eq. (\ref{C3N7}). In Fig. \ref{expectationvalues}(a), the GS exhibits a large value of $\langle\hat{S}_i^z\rangle$ at the edges ($i = 1, 5,$ and $7$). Conversely, in Fig. \ref{expectationvalues}(b), the ES shows the opposite behavior, with values close to zero. This behavior of greater localization for the GS and less for the ES is general for the chains studied in this section, except for the singlet states, for which the expectation value of  $\hat{S}_i^z$ is zero.\\
On the other hand, the correlator between $i$ and $j$ sites defined as $\mathcal{L}_{ij}=\langle\hat{{\bf S}}_i\cdot\hat{{\bf S}}_j\rangle-\langle\hat{{\bf S}}_i\rangle\cdot\langle\hat{{\bf S}}_j\rangle$ gives us useful information about the GS and the ES. The maximum correlation between the edges $|\mathcal{L}_{17}|$ of the previous 3-LSG for the state $\ket{\lambda_{\text{GS}},S=1/2,M=1/2}$ reaches a value of $9\%$ of $S(S+1)$, while for the state $\ket{\lambda_{\text{ES}},S=1/2,M=1/2}$, the value of $|\mathcal{L}_{17}|$ reaches $30\%$ of $S(S+1)$. From this result, we can conclude that the edges in the 3-LSG are more correlated for the ES than for the GS. For the case of an $N=13$ 3-LSG of spin-1/2 MBBs, the correlation between edges is much smaller than that for the $N=7$ case. It is that $|\mathcal{L}_{1(13)}|$ is approximately a $4\%$ of $S(S+1)$ for the $\ket{\lambda_{\text{GS}},S=1/2,M=1/2}$ state and $|\mathcal{L}_{1(13)}|$ is approximately a $17\%$ of $S(S+1)$ for the $\ket{\lambda_{\text{ES}},S=1/2,M=1/2}$ state.  Thus, the correlation between edges is shown to decay more strongly as the length of the legs of the 3-LSG increases, as expected \cite{Kim1998,PhysRevB.55.8894,SBrehmer_1997}.\\
The results presented in this section regarding the double-S degeneracy are general for any spin system described by $H_{\text{BLBQ}}$, provided that the couplings lie within the same phase and the spin species are identical to those of the 3-LSGs.\\

\subsection{Multi-leg and mixed spin graphs}
\label{appA}
\begin{figure*}[htb!]
    \centering
    \includegraphics[width=2\columnwidth]{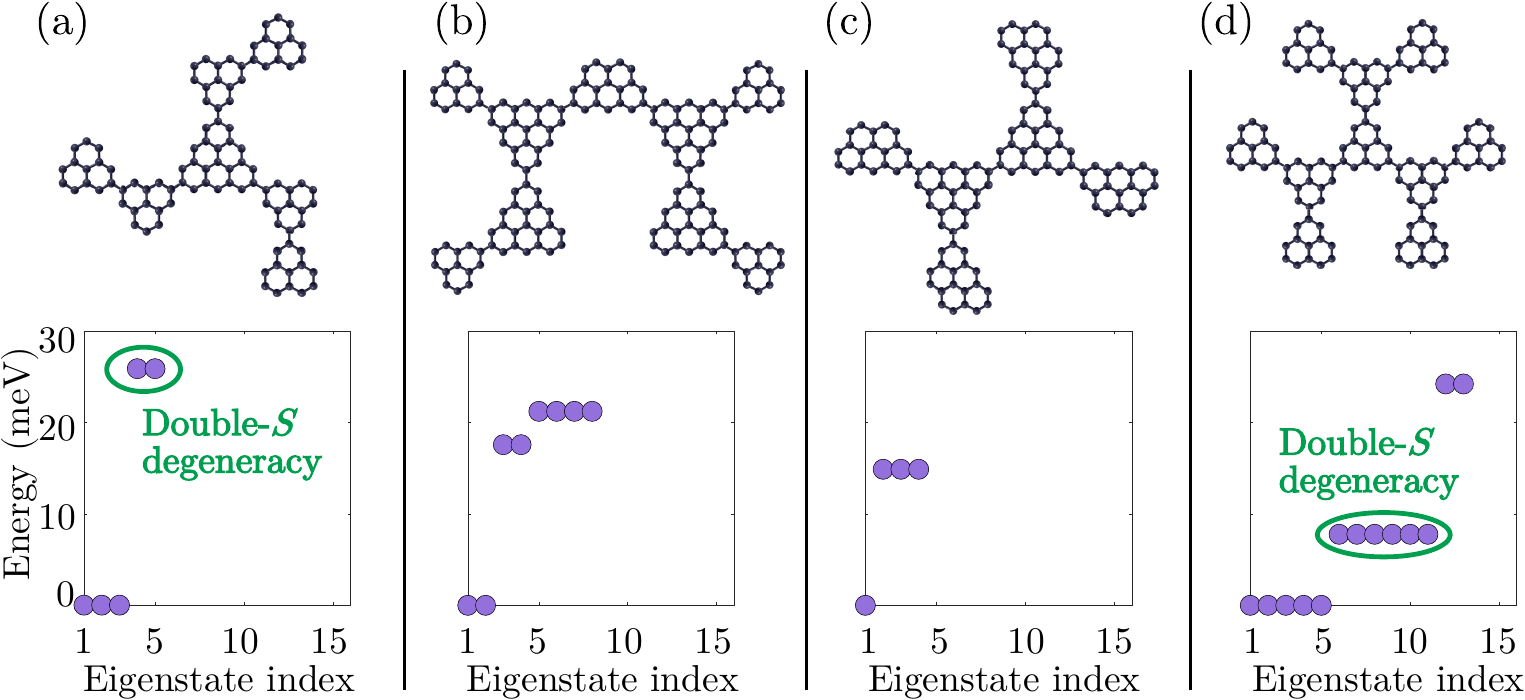}
    \caption{(a) 3-LSG combining S-1T and S-1/2T MBBs. (b-c) Various four-leg spin graphs incorporating S-1T, S-1/2O, and S-1/2T MBBs. (d) Six-leg graph made from S-1/2T. The bottom row displays the GS and the first ES of Eq. (\ref{H_heisenberg}) for each heterospin graph.}
    \label{multileg}
\end{figure*}
Beyond the three LSGs listed in Table \ref{table_multiplets}, an almost limitless number of possibilities emerge when constructing nanostructures by combining different MBBs. In this section, we present selected examples of structures composed of S-1T, S-1/2T, and S-1/2O units. The various combinations of these spin-1 and spin-1/2 MBBs, incorporating different coupling schemes and lengths, render this field a rich playground for exploration, where the properties of both the GS and ESs are highly sensitive to the structural configuration. Some representative examples of these configurations are shown in Fig. \ref{multileg}, while Table \ref{table_multiplets_multi} summarizes the corresponding $S$ values for the GS and ES, along with the energy gaps. The bottom row of Fig. \ref{multileg} displays the low-energy spectrum of the Hamiltonian defined in Eq. (\ref{H_heisenberg}).\\

Fig. \ref{multileg}(a) depict modifications of the 3-LSGs presented in the previous section, where the central triangulene species differs from that of the legs. The 3-LSG shown in Fig. \ref{multileg}(a) exhibits a two-fold degenerate singlet ES, similar to those studied in Table \ref{table_multiplets}. As we have seen for the case of the 3-LSG, this mixed case is also a direct consequence of the $C_{3v}$ symmetry.\\

Figs. \ref{multileg}(b–c) show examples of possible four-leg spin graphs composed of S-1T, S-1/2T, and S-1/2O as MBBs. In these four-legged nanostructures, no double-$S$ degeneracy arises, since the $C_{3v}$ symmetry is absent.
Finally, Fig. \ref{multileg}(d) shows an example of a six-leg spin graph (6-LSG) made from S-1/2Ts. This 6-LSG also posses a $C_3$ swapping symmetry, and notably, its first excited state is a double-$S$ with $S=1$. In contrast, for a similar 6-LSG in which the spin-1/2 MBBs are replaced by S-1Ts the ground state will be a $S=4$ and the first excited state will be a double-$S$ with $S=3$. Larger combinations of MBBs can form nanostructures with an arbitrary number of legs. However, engineered magnetic graphene-based nanostructures can be constructed by focusing on the $C_{3v}$ symmetry. For the S-1T and S-1/2T MBBs, this can be achieved with a number of legs equal to $3\cdot2^{N}$ with $N=0,\,1,\dots$ corresponding to the bipartition of edges at the vertices of the spin graph. Nevertheless, nanostructures with more than six legs may not be of experimental interest. This is because the gap between the GS and the first ES decreases as the number of MBBs increases, while the lifting of the double-$S$ degeneracy due to noise becomes more relevant, as we will discuss in section \ref{appJK}.

 \begin{table}[htb!]
\small
  \caption{$S$ values for the GS and the first ES computed by ED, as well as the energy gap between them, for the spin graphs depicted in Fig. \ref{multileg}. }
\begin{tabular*}{0.5\textwidth}{@{\extracolsep{\fill}}cccc}
    \hline
    System & GS &  ES & Gap (meV) \\ 
    \hline
     (a) & $S=1$ & $S=0$ (two-fold deg.)& 25.9\\   
     (b)&$S=1/2$ &$S=1/2$&  17.6  \\ 
     (c)& $S=0$&$S=1$ &  14.9 \\ 
     (d)&$S=2$&$S=1$ (two-fold deg.)&  7.8\\ 
    \hline
\end{tabular*}
  \label{table_multiplets_multi}
\end{table}

\subsection{Robustness of degeneracies against magnetic anisotropy and exchange coupling noise}
\label{appJK}
\begin{figure}[h]
    \centering
    \includegraphics[width=\columnwidth]{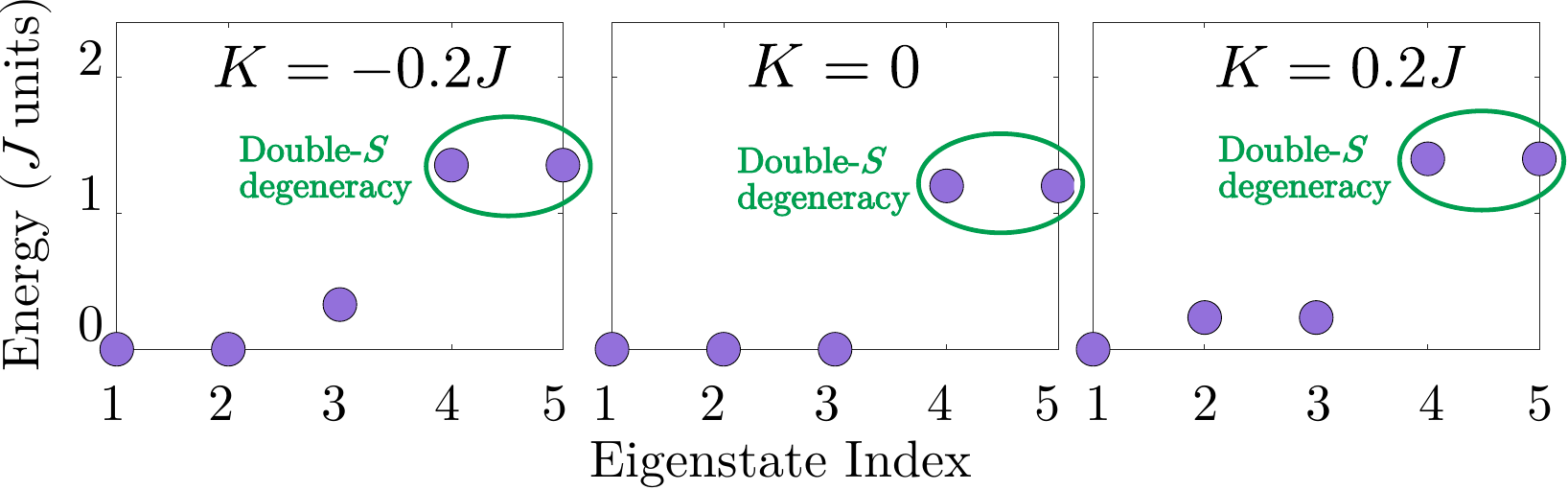}
    \caption{It is shown how the magnetic anisotropy constant affects the energy spectrum of the $N=7$ 3-LSG composed of spin-1 MBBs. A finite $K$ ($K \neq 0$) does not lift the double-$S$ degeneracy of the first excited state, whereas the original triplet is split.
}    
    \label{appendixB-K}
\end{figure}

A fundamental question regarding the degeneracies observed in these 3-LSG systems is whether they might be resolved under realistic experimental setting. To address this, we adopt two approaches applied to the $N=7$ 3-LSG composed by spin-1 MBBs depicted in Fig. \ref{3legchain1}(a). The first approximation consists of adding a magnetic anisotropy term $\hat{H}_{\text{ani}} = K\sum_{i}(\hat{S}_i^z)^2$ \cite{PhysRevB.110.184406,He2021,ref1}  to the Hamiltonian given by Eq. (\ref{H_heisenberg}), where $K \in [-J, J]$ denotes the magnetic anisotropy constant. Including this anisotropy term implies that the quantum number $S$ is no longer good quantum number, but we will focus exclusively on the energy spectrum. This magnetic anisotropy term could be found experimentally by using substrates where the spin-orbit interaction is strong, such as Au(111) \cite{exp_spin1,exp_spin12,exp_olimp}. From a classical point of view, positive (negative) $K$ favors in-plane (out-of-plane) magnetization. Importantly, this anisotropy term does not break the $C_{3v}$ symmetry, which, as we will show below, is responsible for the persistence of the double-$S$ degeneracies. Figure \ref{appendixB-K} shows how the magnetic anisotropy $K$ affects the energy spectrum. A finite 
$K$ $(K\neq0)$ only modifies the original triplet by producing a splitting, while the original double-$S$ ES remains degenerate due to the $C_{3v}$ symmetry.\\

In our second approach we added a random contribution to each coupling $J$ between two MBBs and it can be regarded as simulating spatial distortions arising from the experimental setup. This noise contribution is controlled by a parameter $\eta$, so that the couplings become $J_{ij}^\eta = J(1 + \eta_{ij})$, where $\eta_{ij}$ is a random number uniformly distributed between $-\eta$ and $+\eta$. We define $\{J_{ij}^\eta\}$ as the set of all the exchange constants for a given nanostructure with the noise parameter set to $\eta$. Although the $S$ quantum number remains well-defined, the noise clearly breaks the $C_{3v}$ symmetry, lifting the double-$S$ degeneracy of the first excited state. To quantify this lifting, we calculated the induced gap $\Delta_{\text{ES}}$ in the double-$S$ ES. Table \ref{table_appendix_noise} demonstrates that the induced gap scales linearly with $\eta$. For an $N=7$ 3-LSG composed of spin-1 MBBs, when a noise value of $\eta=0.05$ ($\eta=0.10$) is applied, the induced gap in the ES remains 20 (10) times smaller than the gap between the GS and the first ES as shown in Table \ref{table_multiplets}. However, our results suggest that for larger 3-LSGs, this level of noise may be sufficient to produce a lifted double-$S$ gap of comparable magnitude to the gap between the GS and the first ES.
 \begin{table}[htb!]
\small
  \caption{Induced gap $\Delta_{\text{ES}}^{\text{mean}}$ in the double-$S$ ES, averaged over $10^3$ randomly chosen sets of exchange constants $\{J_{ij}^\eta\}$ with the noise parameter set to $\eta$ for an $N=7$ 3-LSG composed of spin-1 MBBs. The gap between the GS and the first ES is 0.60$J$. }
\begin{tabular*}{0.2\textwidth}{@{\extracolsep{\fill}}cc}
    \hline
    $\eta$ & $\Delta_{\text{ES}}^{\text{mean}}$ ($J$ units) \\ 
    \hline
     0 &0\\ 
     0.05&0.03\\ 
     0.1& 0.06 \\ 
     0.15& 0.09 \\ 
     0.2& 0.12\\ 
    
    \hline
\end{tabular*}
  \label{table_appendix_noise}
\end{table}

\section{Summary and Conclusions} 
\label{Sec_conclusions}
We compute the bilinear and biquadratic exchange constants for various dimer systems composed of different magnetic nanographene structures combining three different first-principles approaches. DFT \cite{PhysRev.136.B864,PhysRev.140.A1133,Soler2002} calculations were used to obtain the electronic Hamiltonian, then the LKAG formalism \cite{Liechtenstein1987,PhysRevB.68.104436,PhysRevB.99.224412} enabled us to extract the exchange constants from Eq. (\ref{H_heisenberg}). Then we employed ED method \cite{PhysRevLett.113.127204} to compute the energy spectrum and the quantum numbers of the eigenstates.
We propose a theoretical realization of alternating spin-1 and spin-1/2 chains composed of different MBBs, such as S-1T, S-1/2T, and S-1/2O. These alternating chains exhibit ferrimagnetism, and both the total quantum spin $S$ and the GS degeneracy scale linearly with the length of the chain.We then characterize the low-energy eigenstates of 3-LSGs of varying lengths ($N = 7$, 10, and 13 MBBs) and compositions (either S-1T or S-1/2T). In the first ES, we identify a double degeneracy in the total spin quantum number $S$. 
 We showed that this unexpected feature of the spectrum arises due to the symmetry of the Hamiltonian with respect to swapping transformations that form a group which is isomorphic to the $C_{3v}$ point group. This double-$S$ degeneracy is protected by the $C_{3v}$ symmetry, which is lost when random noise is introduced into the exchange constants. However, at reasonable levels of noise the lifting of the first ES remains at least an order of magnitude smaller than the gap between the GS and the first ES for an $N=7$ 3-LSG composed of spin-1 MBBs, indicating that the experimental observation of the double-$S$ degeneracy in such systems could be feasible. Moreover, an anisotropy term does not break the $C_{3v}$ symmetry and therefore does not affect
the double-$S$ degeneracy of the first ES. The identification of symmetry-induced degeneracies in 3-LSGs highlights the potential relevance of such symmetries in other quantum systems governed by Hamiltonians with analogous structural properties.\\ These findings open new avenues for the experimental design of custom spin graphs using MBBs. They also lay the groundwork for exploring quantum degeneracies dictated by discrete spatial symmetries in engineered nanostructures. We also presented and characterized some examples of heterospin graphs combining S-1T, S-1/2T, and S-1/2O MBBs. In some cases, the double-$S$ degeneracy in the first ES persists whenever the $C_{3v}$ symmetry is present, as observed in the six-legged spin graphs but not in the four-legged ones. Moreover, since the results presented in this paper rely on ED calculations of the BLBQ Hamiltonian, they can be generalized to physical systems beyond those based on nanographene structures, provided that the spin species and couplings are similar to those considered in this work.

\section*{Acknowledgements}
G. M.-C., A. G.-F. and J. F. have been funded by Ministerio de Ciencia, Innovación y Universidades, Agencia 
Estatal de Investigación, Fondo Europeo de Desarrollo Regional via the grants PGC2018-094783 and PID2022-137078NB-I00,
and by Asturias FICYT under grant AYUD/2021/51185 and by Agencia SEKUENS (Asturias) under grant UONANO IDE/2024/000678 with the support of FEDER funds. G. M.-C. has been supported by 
Programa ``Severo Ochoa'' de Ayudas para la investigación y docencia del Principado de Asturias.
This work was supported by the Ministry of Culture and Innovation and the National Research, Development and Innovation Office within the Quantum Information National Laboratory of Hungary (Grant No. 2022-2.1.1-NL-2022-00004)
and projects K131938, K142179  ADVANCED 149745. We thank the ”Frontline” Research Excellence Programme of the NRDIO, Grant No. KKP133827. This project has received funding from the HUN-REN Hungarian Research Network. This project is supported by the TRILMAX Horizon Europe consortium (Grant No. 101159646).


%

\end{document}